\documentclass[conference]{IEEEtran}

\usepackage{balance}

\usepackage{epsfig}

\IEEEoverridecommandlockouts
\usepackage{graphicx}
\usepackage{algorithmic}
\usepackage{algorithm}
\usepackage{listings}
\usepackage[OT4,T1]{fontenc}
\usepackage[cmex10]{amsmath}
\usepackage{url}
\usepackage{multirow}

\title{In-Bed Person Monitoring \\ Using Thermal Infrared Sensors\thanks{This work has been done by E. Josse, A. Nerborg in the context of their BSc Thesis in Computer Engineering. Authors K. Hernandez-Diaz and F. Alonso-Fernandez thank the Swedish Research Council for funding their research.}}
\author{
\IEEEauthorblockN{ Elias Josse, Amanda Nerborg, Kevin Hernandez-Diaz, Fernando Alonso-Fernandez}
\IEEEauthorblockA{
School of Information Technology (ITE), Halmstad University, Sweden\\
Email: kevin.hernandez-diaz@hh.se, feralo@hh.se}
}

\begin{document}
\maketitle              

\begin{abstract}
The world is expecting an aging population and shortage of healthcare professionals. This poses the problem of providing a safe and dignified life for the elderly.
Technological solutions involving cameras can contribute to safety, comfort and efficient emergency responses, but they are invasive of privacy. We use 'Griddy', a prototype with a Panasonic Grid-EYE, a low-resolution infrared thermopile array sensor, which offers more privacy.
%
%
Mounted over a bed, it can determine if the user is on the bed or not without human interaction. 
For this purpose, two datasets were captured, one (480 images) under constant conditions, and a second one (200 images) under different variations such as use of a duvet, sleeping with a pet, or increased room temperature.
We test three machine learning algorithms: Support Vector Machines (SVM), $k$-Nearest Neighbors ($k$-NN) and Neural Network (NN). With 10-fold cross validation, the highest accuracy in the main dataset is for both SVM and $k$-NN (99\%). 
The results with variable data show a lower reliability under certain circumstances, highlighting the need of extra work
to meet the challenge of variations in the environment.
\end{abstract}

\section{Introduction}
\label{sec:intro}

The proportion of working-age population 
is reducing due to increased life expectancy.
A large number of people born in the baby-boom of the 60s is now starting to be in need of elderly healthcare.
In Sweden, for example, the population beyond 80 years old will increase by 76\% in 2035, in parallel to a shortage of healthcare professionals, partly due to a lack of interest in such education amongst young people \cite{SCB16}.
As a result, monetary resources (work-related taxes) and dedicated workforce will be proportionally less. 
There is also a shortage of elderly residences, which is not expected to be solved easily soon \cite{Boverket19}.
Other countries are facing similar situations too \cite{EU}.

\begin{figure}[htb]
\centering
        \includegraphics[width=0.18\textwidth]{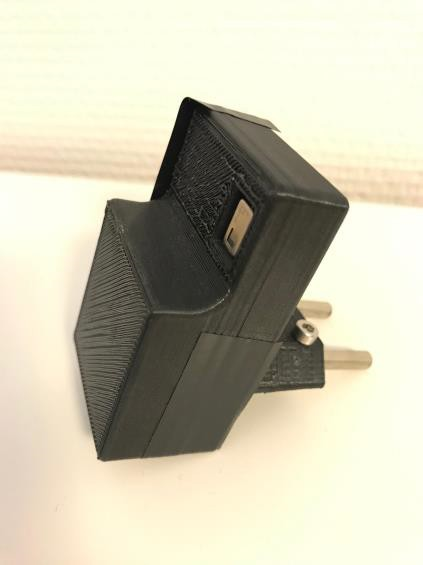}
        \includegraphics[width=0.18\textwidth]{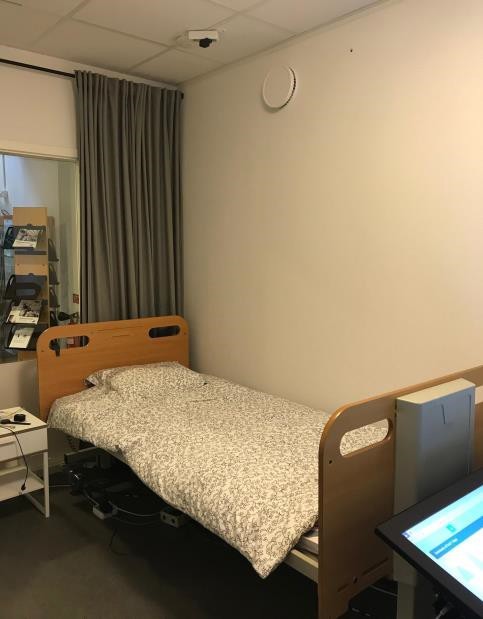}
\caption{Left: Griddy with the Panasonic Grid-EYE infrared array sensor on the front. Right: Bed used in data collection with Griddy in the ceiling.}
\label{fig:griddy}
\end{figure}

This raises the issue of taking care of the elderly in both a humane and economically possible way.
Nowadays, 
people get help in their homes. This can be presumed to be even more common, with people wishing to live autonomously as long as possible.
Home help can be both human and technological, with current solutions containing a mixture that optimizes personnel resources without sacrificing safety.
For example, people can be monitored with sensors at home, specially at night.
Staff can get information of whether the person is in bed or not, receiving alerts if e.g. the person leaves the bed too often or for a long time, or simply if just is sleeping poorly.
%
%
%

Cameras in visible range need sufficient light to work. They also pose privacy issues, since people can be recognized.
One way is to use low resolution cameras, making recognition difficult.
The Grid-EYE sensor from Panasonic offers an 8$\times$8 thermal image that also can operate in darkness \cite{Panasonic}. The low amount of pixels offers more privacy. However, it would be impossible for the personnel to monitor by watching, both because of the low resolution, and because it would demand constant attention, which is infeasible if a reduced amount of staff is supposed to monitor a large amount of people.

Accordingly,  we are interested in predicting if the person is in bed using low-resolution infrared images, so staff can receive alerts and focus only on those who need help.
%
%
%
%
We use a prototype consisting of a Panasonic Grid-EYE sensor with a Bluetooth module (Fig.~\ref{fig:griddy}, left). We call it 'Griddy'. It can be plugged into a 230V socket and collect data at up to 10 frames per second. 
%
%
%
%
The data collection environment is a bedroom of our intelligent home \cite{JensHINT} (Fig.~\ref{fig:griddy}, right).
%
%
Collection is restricted to a single bed and one person at a time.

\begin{figure}[t]
\centering
        \includegraphics[width=0.15\textwidth]{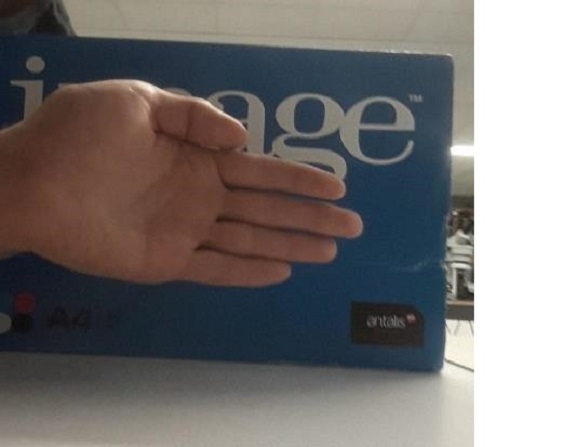}
        \includegraphics[width=0.15\textwidth]{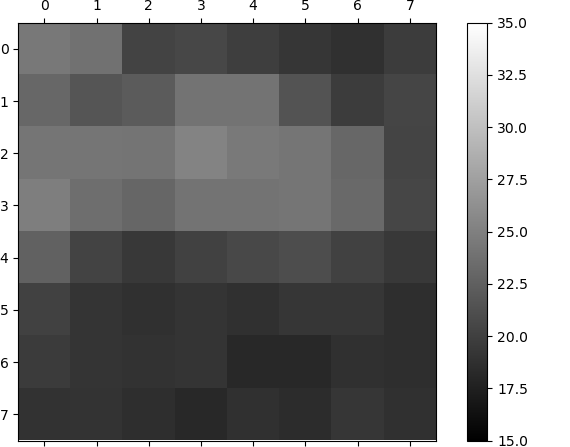}
        \includegraphics[width=0.15\textwidth]{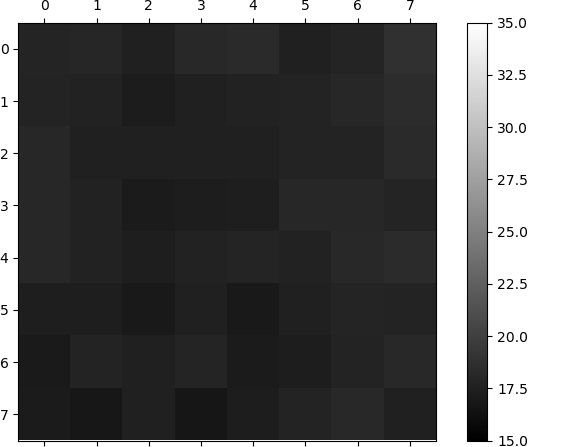}
\caption{Left: hand at 30 cm from the camera. Center: capture with Griddy of the scene on the left. Right: capture without hand in the scene.}
\label{fig:hand}
\end{figure}


\subsection{Related Works}
\label{sec:related_works}

Infrared radiation (IR) is electromagnetic radiation covering a wavelength longer than visible but shorter than millimeters. IR detectors can be categorized into thermal (responsive to heat) and photon (to light) \cite{Rogalski20,Shetty17}. 
Passive sensors just capture radiations from objects and are used e.g. to trigger alarms or handle lights automatically.
Active sensors transmit and collect the response of radiated elements, allowing to distinguish and track objects \cite{Rogalski20,Shetty17}.
%
%
%
The sensor that we use is of active type.

Low-resolution thermal sensors have been widely used for human detection indoors. Their advantages are low price, privacy preservation, and operation with low light \cite{Trofimova17,Shetty17}. The Panasonic Grid-EYE has been used in several works. In \cite{Trofimova17}, 
to improve accuracy by considering temperature variations from other sources than humans. In \cite{Shetty17}, to detect and track moving humans.
In \cite{Chen18}, the authors combined Grid-EYE with an ultrasonic sensor (HC-SR04) for fall-detection. 
The algorithm used, SVM, had the task of differentiating between a fall and another event. 
In \cite{Pontes17}, they used the OMRON D6T-44L thermal sensor of 4$\times$4 pixels installed in a bedroom ceiling to recognize body pose and presence, for which they used decision trees. The results indicated that accounting with data with sufficient diversity was of great importance for a good performance. This motivates us to capture data under several environment variations.
In \cite{Liu10}, they developed a fall detection system using $k$-NN to classify the body posture. To differentiate between fall and lying down, time differences were used. The body silhouette was used for more privacy. 

\begin{figure}[t]
\centering
        \includegraphics[width=0.42\textwidth]{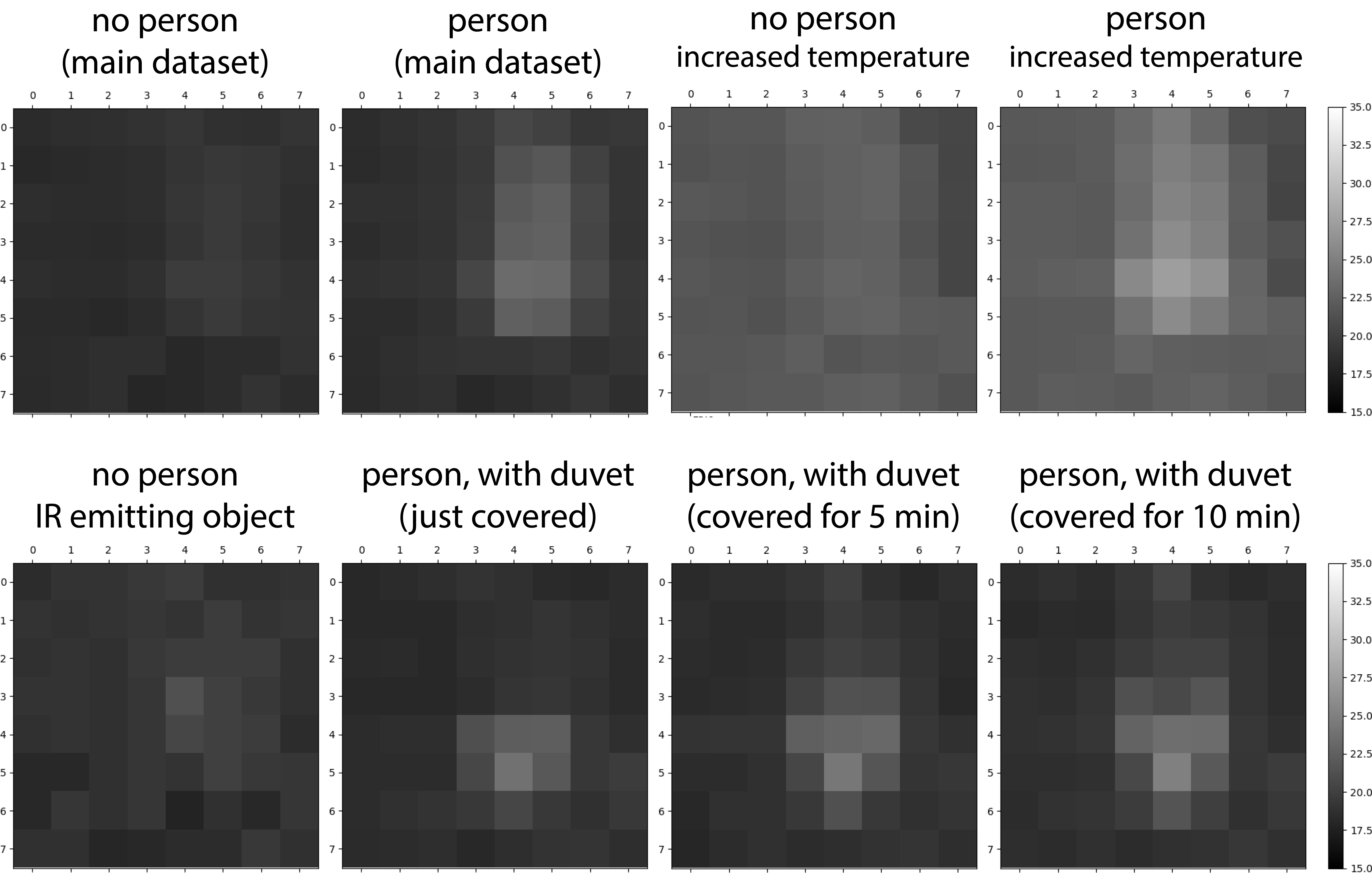}
\caption{Mean temperature (per pixel) of several classes.}
\label{fig:mean-temps}
\end{figure}

\begin{figure}[t]
\centering
        \includegraphics[width=0.22\textwidth]{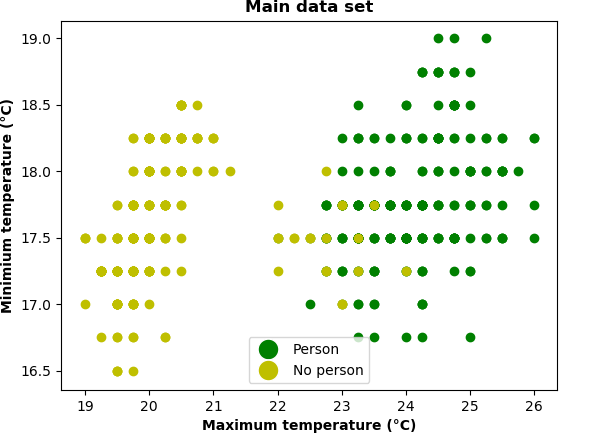}
        \includegraphics[width=0.22\textwidth]{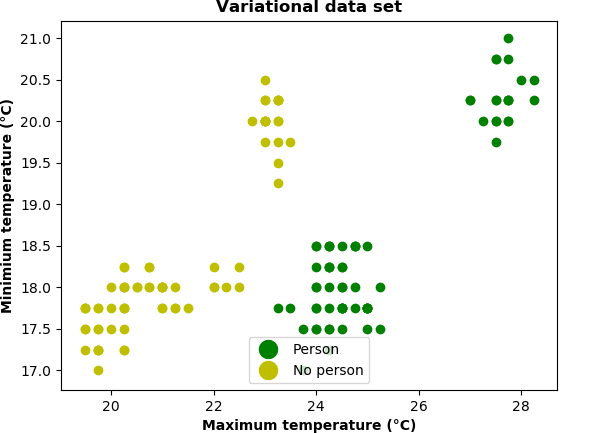}
\caption{Scatter plots (maximum and minimum values per image) of the databases for the person/no person images.}
\label{fig:scatter-temps}
\end{figure}

\section{Methodology}
\label{sec:methods}

\subsection{Software Components}
\label{sec:software}

Three machine learning algorithms, Support Vector Machines (SVM) \cite{Cortes95}, $k$-Nearest Neighbors ($k$-NN) \cite{Dudani76}, and Neural Networks (NN) \cite{Alpaydin10} are used for classification.
SVM finds an hyperplane in the feature space that maximizes the margin (minimum distance between the decision boundary and the closest samples, called support vectors). 
%
%
%
$k$-NN is one of the simplest machine learning algorithms. It computes the distance of unclassified samples to all samples of the training set, and assigns the class that is most represented in the $k$ nearest neighbours.
It is simple, but slow at predicting 
since it has to compute the distance to the entire training set. 
NN are modelled after the human brain. They have several layers with a number of neurons 
per layer. The input is passed through one or more layers (called hidden layers), where the neurons of each layer weight and combine the input of the previous layer, and the output is then passed onto the next layer. 
The weights of the hidden layers are learned during training, allowing to learn patterns of the input data and model classification functions that can predict the label of a given sample.
NN has the ability of handling a large number of training samples, and it is extremely fast in prediction. but 
it is slow to train.

The code of this work is in Python using the Scikit library.
For each algorithm, the most suitable parameters must be found.
Four kernels are used for SVM: linear, polynomial, Radial Basis Function (RBF) and sigmoid.
A range of different $k$ ($k$-NN) and number of neurons (NN) are tested to find the optimum ones for our task.
The number of neurons in the hidden layers are usually between the size of the input layer (input dimensionality) and the size of the output layer (number of classes).
Two hidden layers usually allow to model complex problems with many classes, but there is risk of under-fitting with few data, as in our case \cite{Stathakis09}. Thus, we will employ one hidden layer, with 1 to 1024 neurons
%
%
%


\begin{figure*}[htb]
\centering
        \includegraphics[width=0.92\textwidth]{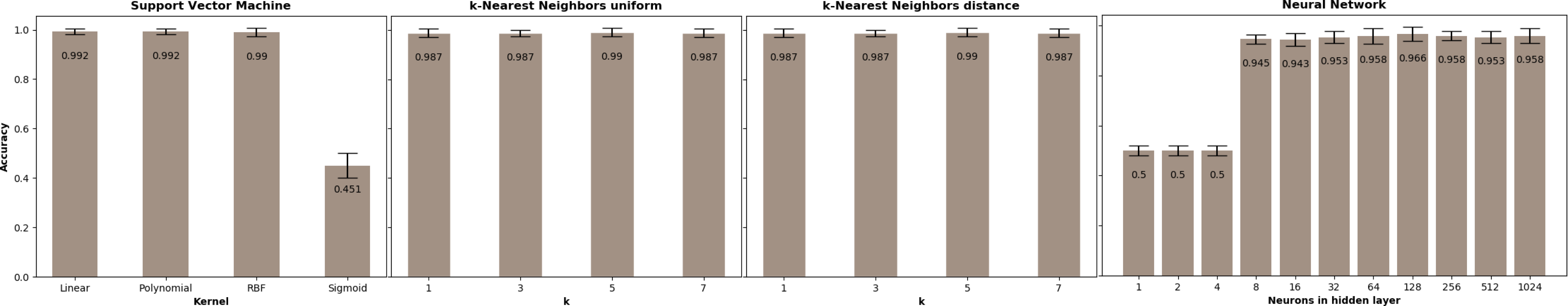}
\caption{Accuracy of the different algorithms on the main dataset using 10-fold cross validation. The standard deviation is shown with an error bar.}
\label{fig:results-main}
\end{figure*}

\subsection{Hardware Components}
\label{sec:hardware}

Grid-EYE, the sensor used, is an active infrared array sensor with 64 thermopile elements arranged in a 8$\times$8 grid. The elements provide a temperature value each. 
The angle of vision is 60 degrees both horizontal and vertical and the distance of use is up to 7 meters. The output range is 20-100 degrees Celsius, rounded to a quarter degree \cite{Panasonic}. Figure~\ref{fig:hand} shows an example of capture of a human hand, with a background surface (cardboard) used to protect from inferences.

\section{Database and Protocol}
\label{sec:db-protocol}

Different datasets have been collected, one referred as the main dataset, and the second one referred as the variational dataset.
Both have been captured at our intelligent home [reference hidden due to double-blind], as shown in Figure~\ref{fig:griddy} (right).
The bed measures 0.9 $\times$ 2 m. Griddy was fixed in the ceiling 2 m over the bed. Thus,
%
%
the captured area at the bed level is 2.3 $\times$ 2.3 m. 
For the label 'no person', 
sometimes a person was present just outside the view area. 
Also, some data was collected from a maximum 30 seconds after a person had been in the bed, while other times it had been several hours.
For the label 'person', six different people were presented. They were asked to vary between different sleeping positions to simulate realistic human positioning in the bed. The test group consisted of both male and female adults with different heights and body types.
The aim was to get as much diversity as possible in body temperature, shape, and position.

The \textit{main} dataset has 480 images (240 'person', 240 'no person').
It was collected during 4 different days across 4 weeks. Data of the two classes were equally distributed over the 4 days. Every day, the collection alternated 20 captures of one class and 20 of the other. The order of collection varied from day to day.
Fig.~\ref{fig:mean-temps} (top left) shows the mean temperature of the images of each class (pixel-wise average of all images).
Fig.~\ref{fig:scatter-temps} (left) shows the scatter plot of the maximum and minimum values per image of each class.
Acquisition across different days and people results in small differences, but the classes are grouped in two clusters, which is expected given the difference in temperature between a person and the room. The data also appears to be linearly separable.

The \textit{variational} dataset incorporates three variations: higher room temperature, a hot non-human object present, and a duvet covering the person. These were chosen because they are expected to occur frequently. 
For each one, 20 images were collected with a person and 20 without.
To increase room temperature, a portable radiator was used to go from the standard 20-21 to 24-25 degrees Celsius.
Fig.~\ref{fig:mean-temps} (top right) shows the mean temperature of the images under this condition.
For the second variation, a 
bottle with warm water at 37 C was used to simulate a small pet. The bottle was placed on top of the bed in various positions. %
Fig.~\ref{fig:mean-temps} (bottom left) shows the mean temperature of the images with the water bottle present and no person.
The last variation was a duvet. The person was covered up to the neck. Ten images with 'person' were collected right after the person had gotten into bed and covered. Another ten images were collected after five minutes, and another ten after ten minutes.
Fig.~\ref{fig:mean-temps} (bottom right) shows the mean temperature of each class.
Finally, the scatter plot of the maximum and minimum values per image of the variational dataset is depicted in Figure~\ref{fig:scatter-temps} (right).
The data forms several clusters given the wider range of variations, specially the maximum temperature, which reaches higher values. Still, the classes appear to be linearly separable.
In our examination of the data, the persons appear to heat up the duvet after a while, reaching the same levels than the human body itself. This can be seen 
in the evolution of the three mean images of Fig.~\ref{fig:mean-temps} (bottom right).
The heat of the water bottle is also detectable when there is no person, but its heat pattern and levels are not equal to those of a person. 

\begin{figure}[t]
\centering
        \includegraphics[width=0.42\textwidth]{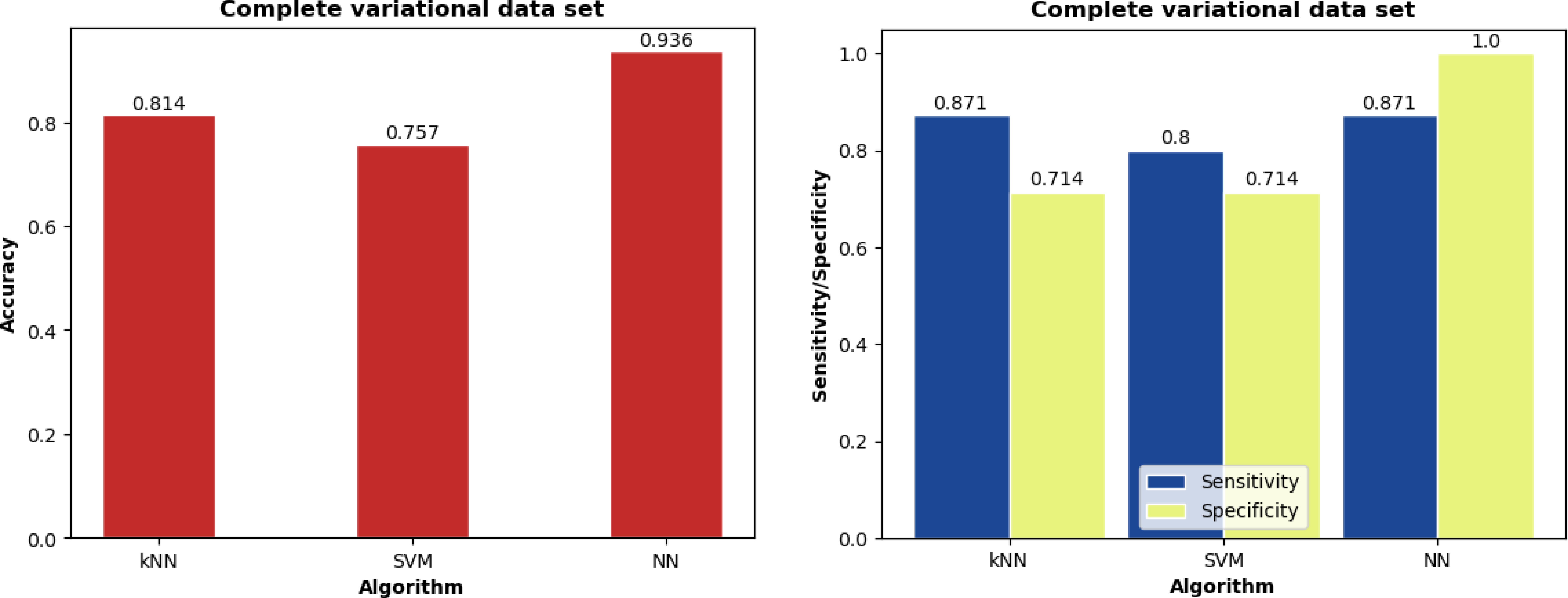}
\caption{Accuracy, sensitivity and specificity (variational dataset).}
\label{fig:results-variational-entire}
\end{figure}

\begin{figure}[t]
\centering
        \includegraphics[width=0.42\textwidth]{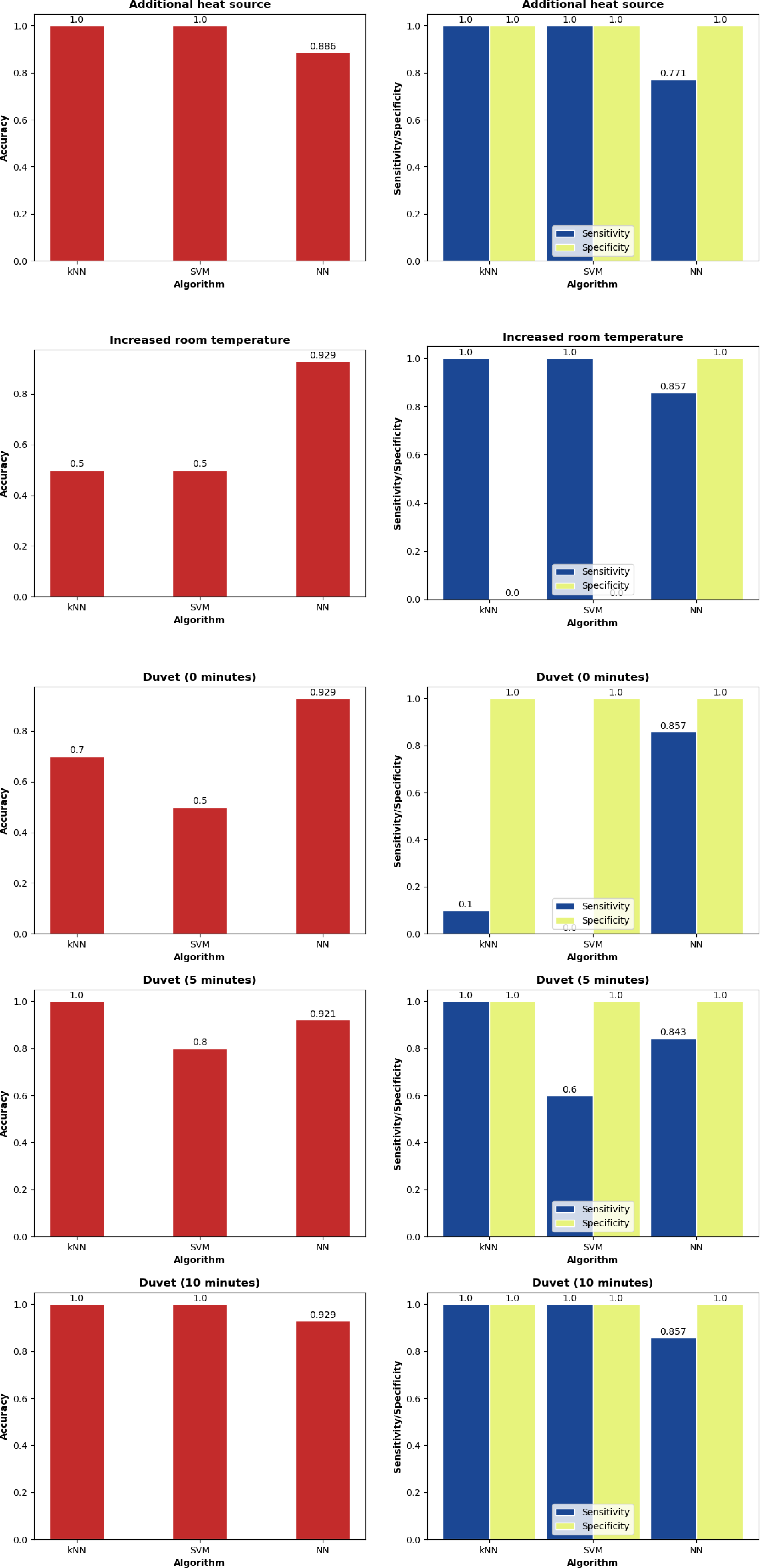}
\caption{Accuracy, sensitivity and specificity (subsets of variational dataset).}
\label{fig:results-variational-parts}
\end{figure}

%
One data sample (input image) consists of a vector of 64 values (number of pixels), which is used as input of the classifiers.
The output labels are the classes person/no person.
%
%
The main data set was divided randomly into 80\% (training) and 20\% (test), with both classes equally represented.
10-fold cross-validation was used due to the fairly small dataset. The same splits were used for all algorithms and all parameter tweaks, so the same folds were always used. 
%
%
The best configuration of the classifiers (found with the test set) were then retrained
on the entire main set, and then evaluated on the variational set.

To compute the success of the predictions, we use the True/False Positive (TP/FP, the system predicts that there is a person in bed, and there is/there is not in reality) and True/False Negative (TN/FN, the system predicts that there is not a person in bed, and there is not/there is in reality). Both FP/FN are system errors, but
%
%
the consequence of each is different \cite{Burkov19}. For example, a FP 
could lead to a potential critical situation that goes unnoticed (the person has left the bed). A FN, 
on the other hand, would send an alarm to an operator, when there is no issue in reality.
As performance metrics, we employ accuracy, sensitivity and specificity \cite{Altman13}.
The accuracy quantifies the right predictions in proportion to the total amount of predictions done, regardless of the actual class, computed as
$Accuracy = {({TP + TN})}/{({TP + TN + FP + FN})}$.
%
%
%
On the other hand, sensitivity describes how well the algorithm predicts positive labels (that the person is in bed), measured as:
$Sensitivity = {({TP})}/{({TP + FN})}$.
%
%
%
Finally, specificity describes how well the algorithm predicts negative labels (that the person is not in bed), as:
$Specificity = {({TN})}/{({TN + FP})}$.
%

\section{Experiments and Results}
\label{sec:experiments}

The algorithms are first evaluated on the main dataset to find the best settings (Fig.~\ref{fig:results-main}).
With SVM, linear, polynomial, Radial Basis Function (RBF) and sigmoid kernels are used.
For $k$-NN, we test $k$=1, 3, 5, 7.
We also test two possibilities, one where the $k$ closest neighbours contribute equally to the decision (second column, 'uniform'), and another where the contribution of each neighbour is weighted by the inverse of its distance to the test sample (third column, 'distance'). The uniform approach gives the same importance to each neighbour, while in the distance approach, the closest neighbours are given more importance.
With the NN, the number of neurons of the hidden layer is varied from 1 to 1024. 

The results show that a high accuracy in general can be obtained with any classifier.
The best result with SVM (99\% accuracy) is given by several kernels, so for subsequent experiments, we use the linear kernel, since more complex kernels do not show a better accuracy.
For $k$-NN, the accuracy with one neighbour ($k$=1) is already very high (99\%). From these results, the value of $k$ with the variational dataset will be $k$=1. The experiments between uniform and weighted distances do not show differences either, very likely because the results with $k$=1 are already nearly to 100\% accuracy, so weighting neighbours with their distance does not provide additional gains.
Lastly, accuracy with the NN is maximized when 128 neurons are employed (97\%), which is the configuration retained for further experiments. Changing the neurons to more or less than 128 has a slight impact, with the accuracy being 94-96\%.
With less than 8 neurons, the accuracy falls dramatically, being equivalent to tossing a coin (50\%).

The algorithms with their best settings are then compared with regards to accuracy, sensitivity and specificity on the variational dataset.
Figure~\ref{fig:results-variational-entire} shows the results.
The best accuracy is with NN, which is a little below the accuracy on the main dataset (94 vs. 97\%).
The other two classifiers showed 99\% accuracy on the main dataset, but here they go down to 81\% ($k$-NN) and 76\% (SVM)
With regards to the other metrics, the NN fails in predicting the positive labels (when the person is in bed), with a sensitivity of 87\%.
When there is no person in bed, it shows a 100\% success (specificity).
This is good in principle, because the classifier never misses to detect that a person has left the bed.
However, on some occasions, there would be false alarms (i.e. the person is really in bed).
With the other classifiers, the behaviour is opposite. They have better sensitivity than specificity, which in principle is not as desirable in our scenarios.

We further report the accuracy on subsets of the variational dataset, according to the different variations.
Figure~\ref{fig:results-variational-parts} (first row) shows the results with an additional heat source in the form of a filled water bottle, resulting in 100\% accuracy with both $k$-NN and SVM. As seen earlier, these two classifiers were severely affected when passing from the main dataset to the variational dataset, but it seems that this is not the perturbation that produced such change.
The NN, on the other hand, sees its accuracy reduced to 89\%, so it seems to be more affected by this perturbation.
Still the NN has 100\% specificity, being the sensitivity the metric that is affected. From this point of view, the capacity of the NN to detect when there is no person in bed goes untouched, but it sees increased its number of false alarms.

The results with increased room temperature from 21-22 to 24-25 degrees Celsius are given in Figure~\ref{fig:results-variational-parts} (second row).
Interestingly, $k$-NN and SVM are severely affected with this perturbation, while in the previous one (temperature increase in only a small region), they performed very well.
Its capacity to detect that the person is not in bed (specificity) falls to zero
The NN is not as affected as with the previous perturbation, with its accuracy recovered to 93\%, and its specificity intact.

Finally, the results with a duvet put at different moments are given in the last three rows of Figure~\ref{fig:results-variational-parts}.
The performance of the NN (in any of its metric) is independent on the time passed. Its accuracy, sensitivity and specificity is similar to the previous perturbation, and relatively close to the metrics on the entire variational dataset (Figure~\ref{fig:results-variational-entire}). This leaves the NN as the best classifier overall, since it appears to be resilient to the majority of perturbations introduced, the only exception being the use of an additional heat source to simulate a pet.
With regards to the other two classifiers, we can observe that its accuracy improves as the time with the duvet on increases. With the person just covered (0 minutes ago), they are mostly useless. However, when the person has spent several minutes covered with the duvet, they are capable of obtaining a 100\% accuracy (specially $k$-NN, which achieves that result earlier).

\section{Discussion}

Technical solutions that contribute to safety, comfort and quick help when needed are essential.
The goal of this work is to develop a system that can detect if a bed is occupied or not with an infrared thermal camera placed on the ceiling over the bed.
The camera captures images of just 8$\times$8 pixels, which we demonstrate to be sufficient for our purposes, while ensuring that it is not possible to visually distinguish people.
This can provide a solution for example to monitor elderly persons on the bed.
The person can be monitored with little human interaction, bringing attention of the staff only when there is a potentially dangerous situation, specially at night. %
This would allow a more effective distribution of resources, for example of car rides to the elderly's home.

We have trained and evaluated three different classifiers, namely Support Vector Machines (SVM), $k$-Nearest Neighbors ($k$-NN) and Neural Networks (NN).
They have been compared in terms of accuracy (percentage of correct predictions), sensitivity (percentage of correct positive predictions, i.e. person in bed) and specificity (percentage of correct negative predictions, i.e. no person in bed).
Overall, the three algorithms behave with a similar high accuracy (97-99\%) when trained and tested on the same data conditions.
To test the robustness of the system, we have also introduced variations that can be expected in reality, such as a pet sleeping in the bed alongside the person, changes in room temperature, or the person being covered with a duvet after 0, 5 and 10 minutes.
The NN shows the best performance overall, being highly resilient to the majority of perturbations, while keeping a specificity of 100\%. It means that it is capable of detecting with high accuracy when there is no person in bed under a wide range of perturbations, which is desirable in our scenarios.
That the sensitivity is not 100\% means that there will be false positives (the person is bed but the system says that is not). However, it is better in principle that the system sometimes wrongfully predicts that the bed is empty (raising alarms that turns out to be false), rather than giving false assurance that the person is in bed.
In our case, the sensitivity of the NN is above 84\% in the majority of situations.
On the other hand, the other two algorithms are highly sensitive to some perturbations, for example increased room temperature, or when a person has just been covered with a duvet. In these cases, their sensitivity or specificity falls to zero (depending on the perturbation).
Conversely, with other perturbations, SVM and $k$-NN show 100\% accuracy, which is higher than the corresponding NN accuracy (92-93\%).

%

Such different and opposite behaviour of the classifiers suggests that some sort of classifier combination can be beneficial to cope with image variations, specially with larger databases sizes. 
As future work, we also plan to expand further the variations of the database. Collection has been constrained to only one person maximum in the image. The person, when present, was always laying on the bed, with no others such as sitting up or standing considered. 
For some users, it might be the case that a pet is sleeping in the bed alongside the owner. In this case it is of great importance that the system does not wrongfully determine the bed as occupied in the event where the user has left but the pet stayed. 
A more in-depth analysis is thus needed to determine which kinds of pets and of what sizes the system can handle.
Our experiments also show that some classifiers struggle with room temperature variations. We have tested 20-21 and 24-25 degrees Celsius, but improvements need to be done in this regard, including lower and higher temperatures in the experimentation.

It was not possible to visually identify people by looking at the images, or to guess the gender, but it does not mean that is not technologically impossible. It would also not be entirely impossible to determine what activities are taking place in the field of view of the camera.
The collected data does not need any manual check, so there is not need to store it over time. Discarding it after interpretation would provide privacy to the user in this regard.
The position of the sensor, looking at the bed from the ceiling, may be also controversial. 
One solution could be to have the sensor on the wall, facing the bed horizontally instead.
Another solution could be to cover the room except the bed. If the room is found to be unoccupied, the conclusion would be that the person is on the bed (presuming that the person has not left the home, controlled for example with opportune sensors in the front door).

%

\bibliographystyle{myIEEEtran}



\end{document}